\documentclass[10pt,journal]{IEEEtran}
\usepackage{epsfig,color}
\usepackage{graphicx}
\usepackage{amsbsy}
\usepackage{amsmath}
\usepackage{amssymb}
\usepackage{euscript}
\usepackage{color}

\newcommand {\Define} {\stackrel {\Delta} {=}  }

\begin{document}
\title{Constant-Envelope Precoding with Time-Variation Constraint on the Transmitted Phase Angles} \author{\IEEEauthorblockN{Sudarshan Mukherjee and Saif Khan Mohammed
  } \thanks{Sudarshan Mukherjee and Saif Khan Mohammed are with the Dept. of Electrical
    Engineering, Indian Institute of Technology (I.I.T.), Delhi, India. S. K. Mohammed is also associated
    with the Bharti School of Telecommunication Technology and Management (BSTTM), I.I.T. Delhi. 
    This work was supported by EMR funding from the Science and Engineering Research Board (SERB),
Department of Science and Technology (DST), Government of India.}}
%\onecolumn
\maketitle
%\singlespacing
\begin{abstract}
  We consider downlink precoding in a frequency-selective multi-user
  massive MIMO system with highly efficient but non-linear power
  amplifiers at the base station (BS). A low-complexity precoding
  algorithm is proposed, which generates constant-envelope (CE)
  transmit signals for each BS antenna. To avoid large variations in the phase angle transmitted from each antenna,
  the difference of the phase angles transmitted in consecutive channel uses is limited to $[-\alpha \pi \,,\, \alpha \pi]$ for a fixed $0 < \alpha \leq 1$.
To achieve a desired per-user information rate, the extra total transmit power required
 under the time variation constraint when compared to the special case of no time variation constraint (i.e., $\alpha=1$), is {\em small} 
for many practical values of $\alpha$. In a i.i.d. Rayleigh fading channel with $80$ BS antennas, $5$ single-antenna users and a desired per-user information rate of $1$ bit-per-channel-use, the extra total transmit power required is less than $2.0$ dB when $\alpha = 1/2$. 
\end{abstract}
\vspace{-5mm}
\section{Introduction}\label{sec:SysModel}
In massive MIMO systems a base station
(BS) with a large number of antennas ($N$, a few hundreds)
communicates with several user terminals ($M$, a few tens) on the same
time-frequency resource \cite{SPM-paper, NextGen}.  There has been recent
interest in massive MIMO systems due to their ability to increase
spectral and energy efficiency even with very low-complexity multi-user
detection and precoding \cite{TM,HNQ,SAIF}. However, physically building
cost-effective and energy-efficient large arrays is a challenge.
Specifically in the downlink, the power amplifiers (PAs) used in the
BS should be highly power-efficient. Due to the trade-off
between the efficiency and linearity of the PA \cite{cripps},
highly efficient but non-linear PAs must be used. The efficiency
of the PA is related to the amount of backoff necessitated (to reduce non-linear distortion) by the
peak to average ratio of the input waveform. For minimum
backoff and hence maximum efficiency, the input waveform should have a constant
or nearly {\em constant envelope} (CE).\footnote{\footnotesize{That is, the discrete-time complex baseband signal
transmitted from each BS antenna has a constant magnitude
irrespective of the channel gains and the information symbols to be
communicated.}}

With this motivation, in \cite{prev_work} we had proposed a CE precoding
algorithm for the frequency-flat multi-user MIMO broadcast channel,
which was then extended to frequency-selective channels in \cite{prev_work2}.
With $N$ sufficiently larger than $M$ and i.i.d.\ Rayleigh fading,
numerical studies done in both these papers revealed that
in order to achieve a desired per-user ergodic information rate,
the proposed CE precoding algorithm needed only about $1.0-1.5$ dB extra
total transmit power compared to that required under the less stringent
and commonly used total average transmit power constraint. It was also
observed that even under a stringent per-antenna CE constraint, an
$O(N)$ array gain is achievable, i.e., with every doubling in the
number of BS antennas the total transmit power can be reduced by $3$
dB while maintaining a fixed information rate to each user (assuming that
the number of users is fixed).

However, in the CE precoding algorithm proposed in \cite{prev_work, prev_work2} the phase angle of the
complex baseband signal transmitted from each BS antenna is unconstrained (i.e.,
it's principal value lies in $(-\pi \,,\, \pi]$), and therefore it is possible that the phase could
vary very fast between consecutive channel uses. A phase variation of $180^{\circ}$ (or more) between
consecutive channel uses will result in zero crossings in the baseband signal, which with practical PAs could lead to
distortion in the transmitted signal.
In this paper, we address this problem by proposing a CE precoding algorithm
with an additional constraint that the difference of the phase angle transmitted in consecutive channel uses be limited
to the interval $[-\alpha \pi \,,\, \alpha \pi]$ for a fixed $0 < \alpha \leq 1$ (the special case of $\alpha =1$
was considered in \cite{prev_work2}). It is shown that the complexity of the proposed CE algorithm is
independent of $\alpha$ and is the same as the algorithm proposed in \cite{prev_work2}. Numerical studies
on the i.i.d. Rayleigh fading channel suggest that an $O(N)$ array gain is achieved even under the additional phase angle variation
constraint.
To achieve a desired per-user information rate, the extra total transmit power required
under the time variation constraint when compared to the special case of no time variation constraint (i.e., $\alpha=1$), is {\em small} 
when $\alpha$ is close to $1$ and $N \gg M$.
For example, with $\alpha = 1/2$, $N = 80$, $M=5$ single-antenna users and a desired per-user rate of $1$ bit-per-channel-use (bpcu), the magnitude of the phase variation is limited to $\alpha \pi = 90^{\circ}$ and the extra transmit power required
is less than $2$ dB.
\section{System Model and CE Precoding}
\label{sec-CEprev}
In the previous works and also in this paper, without loss of generality we assume single-antenna users.\footnote{\footnotesize{When a user terminal (UT) has multiple antennas, the proposed algorithm can still be applied by treating each antenna at the UT as a separate user.}}
It is assumed that the BS has knowledge of the channel vector to each user.\footnote{\footnotesize{In a massive MIMO system ($N \gg M$), the amount of time-bandwidth resource required for channel estimation at the BS in the uplink is proportional to $M$, while in the downlink it is proportional to $N$. Since $N \gg M$, it is suggested that massive MIMO systems would operate in a time division duplexed (TDD) mode, so that downlink CSI can be estimated from the CSI acquired in the uplink through uplink training \cite{SPM-paper, NextGen, TM}.}}
The complex baseband constant envelope signal transmitted from the $i$-th BS antenna at time $t$ is of the form
{
\vspace{-1mm}
\begin{eqnarray}
\label{const_env_eqn}
x_i[t] & = & \sqrt{\frac{P_T}{N}} \, e^{j \theta_i[t]} \,\,\,,\,\,\, i = 1,2,\cdots,N,
\end{eqnarray}
} where $j \Define \sqrt{-1}$, $P_T$ is the total power transmitted
from the $N$ BS antennas and $\theta_i[t] \in [-\pi \,,\, \pi)$ is the
phase of the CE signal transmitted from the $i$-th BS antenna at time
$t$.  The equivalent discrete-time complex baseband channel between
the $i$-th BS antenna and the $k$-th user (having a single-antenna) has a finite impulse response
of length $L$ samples, denoted by $(h_{k,i}[0]\,,\, h_{k,i}[1]
\,,\, \cdots \,,\, h_{k,i}[L-1])$.  The signal received at the $k$-th
user ($k=1,2,\cdots,M$) at time $t$ is given by {
\vspace{-2mm}
\begin{eqnarray}
\label{recvk_eqn}
y_k[t] & = & \sqrt{\frac{P_T}{N}} \,\, \sum_{i=1}^N \, \sum_{l=0}^{L-1} \, h_{k,i}[l] e^{j \theta_i[t-l]} \,\,+\,\, w_k[t] \,,\,
\end{eqnarray}
}
where $w_k[t] \sim {\mathcal C}{\mathcal N}(0,\sigma^2)$ is the AWGN at the $k$-th user at time $t$ (AWGN is i.i.d.
across time and across the users). For the sake of brevity let us denote the vector of phase angles transmitted at time instance $t$ by $\Theta[t] = (\theta_1[t], \cdots, \theta_N[t])$.
\begin{figure*}
\vspace{-8mm}
{
\small
\begin{eqnarray}
\label{NLS_joint_eqn}
(\Theta^{u}[1],\cdots,\Theta^{u}[T])   &   =    &
\arg \hspace{-5mm} \min_{\substack{\Theta[t] \in [-\pi,\pi)^N  \\ t=1,\ldots,T}} f(\theta_1[1], \cdots, \theta_N[1], \cdots, \theta_1[T], \cdots, \theta_N[T]) \nonumber \\
\mbox{where} \,\,  f(\theta_1[1], \cdots, \theta_N[1], \cdots,\theta_1[T], \cdots, \theta_N[T]) 
 & \Define & \sum_{t=1}^T \sum_{k=1}^M {\Bigg \vert} \frac{ \sum_{i=1}^N \, \sum_{l=0}^{L-1} \, h_{k,i}[l] e^{j \theta_i[t-l]}   }{\sqrt{N}} -  \sqrt{E_k} u_k[t]   {\Bigg \vert}^2.
\end{eqnarray}
}
\normalsize
\vspace{-4mm}
\end{figure*}
In the following we briefly summarize the CE precoding algorithm proposed in \cite{prev_work2}. 
Suppose that, at time instances $t=1,2,...,T$ we are interested in communicating the information symbol $\sqrt{E_k} u_k[t] \in {\mathcal U}_k \subset {\mathbb C}$ to the $k$-th user. Let ${\mathbb E}[\vert u_k[t] \vert^2] = 1 \,,\, k=1,\cdots,M$. Also, let ${\bf u}[t] = (\sqrt{E_1} u_1[t], \cdots , \sqrt{E_M} u_M[t]) \, \in \, {\mathcal U}_1 \times \cdots \times {\mathcal U}_M$ be the vector of information symbols to be communicated at time $t$.
In \cite{prev_work2}, we had proposed an algorithm for finding the transmit phase angles $\theta_i(t) \,,\, i=1,2,\ldots,N \,,\, t=1,2,\ldots,T$ in such a way that the received noise-free signal at each user is almost the same as the information symbol intended for that user, i.e., $\sqrt{{P_T/N}} \,\, \sum_{i=1}^N \, \sum_{l=0}^{L-1} \, h_{k,i}[l] e^{j \theta_i[t-l]}  \approx \sqrt{P_T} \sqrt{E_k} u_k[t] \,\,,\,\, \forall \, k=1,2,\ldots,M \,,\, t=1,2,\ldots,T$.

In \cite{prev_work2} we find the transmit phase angles as a solution to the optimization problem in (\ref{NLS_joint_eqn}), where $\Theta^{u}[t] = (\theta_1^u[t], \cdots, \theta_N^u[t]) \,,\, t=1,\ldots,T$ denotes the vectors of transmit phase angles, for the given information symbol vectors ${\bf u}[t] \,,\, t=1,\ldots, T$. The main idea in (\ref{NLS_joint_eqn}) is to choose the transmit phase angles in a way so as to minimize the energy of the difference between the received noise-free signal and the intended information symbol for all users.
Note that the objective function $f(\cdots)$ in (\ref{NLS_joint_eqn}) is a function of $NT$ variables ($N$ phase angles transmitted at $T$ time instances). Finding an exact solution to the problem in (\ref{NLS_joint_eqn}) is prohibitively complex, and therefore in \cite{prev_work2} we had proposed a low-complexity near-optimal solution to (\ref{NLS_joint_eqn}).
The CE precoding idea is primarily based on our previous work in \cite{prev_work} (for frequency-flat channels) where we had analytically shown that for a broad class of frequency-flat channels
(including i.i.d. fading), for a fixed $M$ and fixed symbol energy levels ($E_1, \cdots, E_M$), by having a sufficiently large
$N \gg M$ it is always possible to choose the transmit phase angles in such a way that the received signals at the users are arbitrarily close to the desired information symbols. 

\section{CE Precoding with Constrained Time Variation of Transmit Phase Angles} 
\label{sec-CE}
Note that for the CE precoding method, the transmit phase angles can take any value in the interval $[-\pi \,,\, \pi)$ (see (\ref{NLS_joint_eqn})). Therefore it is possible that between consecutive time instances, the phase angle transmitted from a BS antenna could change by a large magnitude, which will distort the transmit signal at the output of the PA. To address this issue, in this paper we propose a CE precoder where for each BS antenna the difference between the phase angles transmitted in consecutive time instances is constrained to lie in the interval $[- \alpha \pi \,,\, \alpha \pi]$ for a given $0 < \alpha \leq 1$, i.e.,
$\vert \theta_i[t]  \, - \, \theta_i[t-1] \vert \leq \alpha \pi $ for all $t=1, \ldots,T \,\,,\,\, i=1,\ldots,N$.
This constraint ensures that the maximum variation in the transmitted phase angle between consecutive time instances is at most $\alpha \pi$ (e.g., with $\alpha = 1/2$ the maximum phase angle variation is only $90^{\circ}$). In this paper, under the time-variation constraint we propose an optimization problem to find the transmit phase angles for given information symbols $u_k[t] \,,\, k=1,\ldots,M \,,\, t=1,\ldots, T$, as given by (\ref{NLS_joint_eqn2}). 

Exactly solving (\ref{NLS_joint_eqn2}) has prohibitive complexity, and therefore in the following we propose a low-complexity near-optimal solution to (\ref{NLS_joint_eqn2}).    
The essential idea of this low complexity solution is to iteratively optimize $f(\cdots)$ as a function of one variable at a time while fixing the other variables to their previous values.
In one iteration of this low-complexity algorithm, we have $NT$ sub-iterations. In the first sub-iteration we start with $\theta_1[1]$ and minimize $f(\cdots)$ as a function
of $\theta_1[1]$ while keeping the other $(NT - 1)$ variables fixed to their previous values. We then update $\theta_1[1]$ with its optimum value and then move onto the second sub-iteration where we minimize $f(\cdots)$ as a function
of $\theta_2[1]$ while keeping the other variables fixed. In general, in the $(N(q-1) + r)$-th sub-iteration we
minimize $f(\cdots)$ as a function of $\theta_r[q]$ (i.e., the phase angle transmitted from the $r$-th BS antenna in the $q$-th time instance) while keeping the other variables fixed.
\begin{figure*}
{
\small
\begin{eqnarray}
\label{NLS_joint_eqn2}
(\Theta^{u}[1],\cdots,\Theta^{u}[T])   &   =    &
\arg \hspace{-5mm} \min_{\substack{\\ (\Theta[1] , \Theta[2], \cdots, \Theta[T]) \\ \vert \theta_i[t] - \theta_i[t-1] \vert \, \leq \, \alpha \pi \\ i=1,\ldots,N \,,\, t=1,\ldots,T }} \, f(\theta_1[1], \cdots, \theta_N[1], \cdots, \theta_1[T], \cdots, \theta_N[T]).
\end{eqnarray}
}
\normalsize
\vspace{-4mm}
\end{figure*}
\begin{figure*}
{
\small
\begin{eqnarray}
\label{new_eqn_1}
\theta^{\prime}_r[q] & = & \arg \hspace{-10mm} \min_{\substack{ \\ \\ \theta_r[q] \\  \hspace{3mm} \vert \theta_r[q]  \, - \, \theta_r[q-1] \vert \, \leq  \,    \alpha \pi}} \hspace{-16mm} \sum_{t=q}^{\min(T, (q + L -1))} \sum_{k=1}^M {\Bigg \vert}  S_{r,q}(k,t) +  \frac{h_{k,r}[t - q] e^{j \theta_r[q]}}{\sqrt{N}}  {\Bigg \vert}^2  \,\,,\,\,
 \mbox{where} \,\, S_{r,q}(k,t)  \, \Define  \,  {\Big (} \sum \limits_{i=1}^N \hspace{-7mm}  \sum \limits_{\substack{l=0 \,,\, \\  \hspace{6mm} (i,l) \ne (r,(t - q))}}^{L-1}  \hspace{-6mm} \frac{h_{k,i}[l] e^{j \theta_i[t-l]}   }{\sqrt{N}} {\Big )} -  \sqrt{E_k} u_k[t] \nonumber \\ 
& = & \arg \min_{\substack{\theta_r[q] \\ (\theta_r[q]  \, - \, \theta_r[q-1]) \,  \in  \, [-\alpha \pi \,,\, \alpha \pi]}} \, \sum_{t=q}^{\min(T, (q + L -1))} \sum_{k=1}^M {\Bigg \vert}  S_{r,q}(k,t) +  \frac{h_{k,r}[t - q] e^{j \theta_r[q-1]}  \, e^{j (\theta_r[q] - \theta_r[q-1])} }{\sqrt{N}}  {\Bigg \vert}^2  \nonumber \\
& = &  \theta_r[q - 1 ] \, + \, \arg \min_{\substack{\omega \,  \in  \, [-\alpha \pi \,,\, \alpha \pi]}} \, \sum_{t=q}^{\min(T, (q + L -1))} \sum_{k=1}^M {\Bigg \vert}  S_{r,q}(k,t) +  \frac{h_{k,r}[t - q] e^{j \theta_r[q-1]}  \, e^{j \, \omega } }{\sqrt{N}}  {\Bigg \vert}^2  \nonumber \\
& = & \theta_r[q - 1 ] \, + \, \arg \max_{\omega  \in [-\alpha \pi \,,\, \alpha \pi]} \,  \Re {\Bigg (} \, e^{j \omega } \, {\Big \{}  \, - \, \sum_{t=q}^{\min(T, (q + L -1))} \sum_{k=1}^M  \,  h_{k,r}[t - q]  e^{j \theta_r[q-1]} S^{*}_{r,q}(k,t)  {\Big \}} {\Bigg )} \nonumber \\
& = & \theta_r[q - 1 ] \, + \,  \left \{
\begin{array}{cc}
\alpha  \pi  \,,  &  \,\,\,  - \pi \, \leq \, c \, < \, - \alpha \pi  \\
- c  \,,  & \,\,\,   - \alpha \pi  \, \leq  \, c \, < \, \alpha \pi  \\
- \alpha \pi  \,, & \,\,\, \alpha \pi  \, \leq  \, c  \leq \pi
\end{array}
\right.
\,\,,\,\,  \mbox{where} \,\, c  \,  \Define  \, \mbox{ARG} {\Big (}  - \hspace{-4mm}  \sum_{t=q}^{\min(T, (q + L -1))} \sum_{k=1}^M  \,  h_{k,r}[t - q]  e^{j \theta_r[q-1]}  S^{*}_{r,q}(k,t) {\Big )} .
\end{eqnarray}
}
\normalsize
\end{figure*}       
Since the channel is causal and has a memory of $L$ time instances, it follows that in the summation on the right hand side of the definition of $f(\cdots)$ in (\ref{NLS_joint_eqn}), only the terms corresponding to $t=q, (q+1), \cdots, \min(T,q+L-1)$ depend on $\theta_r[q]$.
Given this fact, the minimization of $f(\cdots)$ only w.r.t. $\theta_r[q]$ is given by (\ref{new_eqn_1}).
In (\ref{new_eqn_1}), for any complex number $z$, $\mbox{ARG}(z) \Define \{ \phi \in (-\pi \,,\, \pi] \, | \,  e^{j \phi} = z/\vert z \vert \}$ is the principal value of the phase angle of $z$ and $z^*$ denotes the conjugate of $z$.
From (\ref{new_eqn_1}) it is clear that the new value of $\theta_r[q]$ depends on $S_{r,q}(k,t)$. Note that, for every different $(r,q)$ we need not recalculate $S_{r,q}(k,t)$ explicitly using the sum in the R.H.S. of its definition in (\ref{new_eqn_1}). Instead, $S_{r,q}(k,t)$ can be calculated by subtracting the current value of $h_{k,r}[t - q] e^{j \theta_r[q]}/\sqrt{N}$ (i.e., value at the start of the $(N(q-1) + r)$-th sub-iteration) from the current value of $S(k,t) \Define  \frac{ \sum_{i=1}^N \, \sum_{l=0}^{L-1} \, h_{k,i}[l] e^{j \theta_i[t-l]}   }{\sqrt{N}} -  \sqrt{E_k} u_k[t] $, i.e.
{
\begin{eqnarray}
\label{eqn_233}
S_{r,q}(k,t) & = &  S(k,t) \, -  \, \frac{h_{k,r}[t - q] e^{j \theta_r[q]}}{\sqrt{N}}.
\end{eqnarray}
}
Note that with change in $\theta_r[q]$, we also need to change $S(k,t)$ for all $k=1,\ldots,M \,,\, t=q,\ldots,\min(T, q + L - 1)$.
The modified value of $S(k,t)$ after the $(N(q-1) + r)$-th sub-iteration is given by
{
\begin{eqnarray}
\label{eqn_23}
S^{\prime}(k,t) & = & \frac{ \sum_{i=1}^N \, \sum_{{l=0 \,,\,  (i,l) \ne (r,(t - q))}}^{L-1} \, h_{k,i}[l] e^{j \theta_i[t-l]}   }{\sqrt{N}} \nonumber \\
&  &  \, -  \, \sqrt{E_k} u_k[t]  \, + \, \frac{h_{k,r}[t - q] e^{j \theta^{\prime}_r[q]}}{\sqrt{N}}  \nonumber \\
& = & S(k,t) \, +  \, \frac{h_{k,r}[t - q]}{\sqrt{N}}  {\Big (} e^{j \theta^{\prime}_r[q]}  \, - \, e^{j \theta_r[q]} {\Big )}
\end{eqnarray}
}
where $\theta^{\prime}_r[q]$ is the new updated value of the phase angle to be transmitted from the $r$-th BS antenna at time instance $q$, and is given by (\ref{new_eqn_1}).
After the last sub-iteration of an iteration (i.e., where we update $\theta_N[T]$), we start with the first sub-iteration (where we update $\theta_1[1]$) of the next iteration. It is clear that the value of the objective function $f(\cdots)$ reduces monotonically from one sub-iteration to the next. Numerically, it has been observed that the value of the objective function converges in a few iterations ($\leq 5$) and further iterations lead to little reduction in the value of $f(\cdots)$. Further, the value that the objective function converges to, is observed to be small when $N \gg M$. The complexity of each sub-iteration is $O(ML)$ and is independent of $\alpha$ (see (\ref{new_eqn_1})). Since we update $NT$ phase angles in each iteration, the total complexity of each iteration is $O(NMLT)$. With a fixed number of iterations, the overall complexity of the proposed algorithm is $O(NMLT)$, i.e., a per-channel-use complexity of $O(NML)$, which is the same as that of the algorithm proposed in \cite{prev_work2} to solve (\ref{NLS_joint_eqn}).
\section{Information Theoretic Performance Analysis}
For a given set of information symbol vectors ${\bf u}[t] \,,\,t=1,\cdots,T$,
let ${\widehat \Theta}^u[1],{\widehat \Theta}^u[2],\cdots,{\widehat \Theta}^u[T]$ denote the output phase angles of the proposed iterative CE precoding algorithm (see Section \ref{sec-CE}). Let ${\widehat \theta}_i^u[t]$
be the phase angle transmitted from the $i$-th antenna at time $t$.
The signal received at the $k$-th user is then given by
{
\vspace{-2mm}
\begin{eqnarray}
\label{recvk_eqn_tilde}
y_k[t]
& \hspace{-3mm} = & \hspace{-3mm} \sqrt{P_T} \sqrt{E_k} \, u_k[t] \,+\, \sqrt{P_T} I_k^u[t] \, + \, w_k[t] \nonumber \\
 I_k^u[t] & \hspace{-3mm} \Define & \hspace{-3mm} {\Big (}  \sum_{i=1}^N \, \sum_{l=0}^{L-1} \, \frac{h_{k,i}[l]} {\sqrt{N}} e^{j {\widehat \theta}^u_i[t-l]}  \, - \, \sqrt{E_k}  u_k[t]  {\Big )}
\end{eqnarray}
}
Note that $I_k^u[t]$ behaves like multi-user interference (MUI).
Also, let ${\bf y}_k  \Define (y_k[1], \cdots, y_k[T])^T$,
${\bf u}_k \Define (\sqrt{E_k}u_k[1], \cdots,\sqrt{E_k} u_k[T])^T$, ${\bf I}_k^u \Define (I_k^u[1], \cdots, I_k^u[T])^T$ and ${\bf w}_k \Define (w_k[1], \cdots, w_k[T])^T$.
Let ${\bf H} = \{ h_{k,i}[l] \}$ denote the impulse responses of the channels between the $N$ BS antennas and the $M$ users. For a given ${\bf H}$, an achievable rate for the $k$-th user is given by the mutual information
$I({\bf y}_k \, ; \, {\bf u}_k \,\, | \,\, {\bf H} ) / T $ \cite{CTbook}.
For any arbitrary distribution on ${\bf u}_k$,
it is difficult to compute $I({\bf y}_k \, ; \, {\bf u}_k \,\, | \,\, {\bf H} )$. A lower bound on $I({\bf y}_k \, ; \, {\bf u}_k \,\, | \,\, {\bf H} )/T$ is an achievable information rate for the $k$-th user. Therefore, in the following we derive a lower bound to $I({\bf y}_k \, ; \, {\bf u}_k \,\, | \,\, {\bf H} )/T$
assuming $u_k[t] \,,\, t=1,\cdots, T$ to be i.i.d. ${\mathcal C}{\mathcal N}(0, 1)$ i.e., proper complex Gaussian
having zero mean and unit variance.
\begin{eqnarray}
\vspace{-2mm}
\label{inf_rate_bnd}
\frac{I({\bf y}_k \, ; \, {\bf u}_k \,\, | \,\, {\bf H} )}{T} 
%& = & h ({\bf u}_k) \, - \, h ({\bf u}_k \, | \, {\bf y}_k \,,\, {\bf H})  \nonumber \\
&  \hspace{-4mm} = &  \hspace{-4mm}  \frac{1}{T} {\Big (} h ({\bf u}_k) \, - \, h ({\bf u}_k \, - \, {\bf y}_k/\sqrt{P_T} \, | \, {\bf y}_k \,,\, {\bf H})   {\Big )} \nonumber \\
&  \hspace{-3mm} = &   \hspace{-3mm}   \log_2(\pi e E_k)  \, - \, {\Big (} h( {\bf v}_k \, | \,  {\bf y}_k \,,\, {\bf H})  / T {\Big )}  \nonumber \\
&  \hspace{-3mm}   {(a) \atop \geq}  &  \hspace{-2mm}  \log_2(\pi e E_k)  \, - \, {\Big (}  h( {\bf v}_k \, | \,   {\bf H})  \,  /  \, T  {\Big )}  \nonumber \\
&  \hspace{-3mm}  {(b) \atop \geq}  &  \hspace{-2mm}   \log_2( E_k) \, - \,  {\Big (}  \log_2 ( \vert {\bf R}_v \vert)  \, / \, T {\Big )}  \nonumber \\
{\bf v}_k  & \Define &  {\bf u}_k \,  -  \, {\big (} {\bf y}_k/\sqrt{P_T}  {\big )} 
\end{eqnarray}
where $h(\cdot)$ denotes the differential entropy operator, and the inequality in step (a) is due to the fact that conditioning reduces entropy \cite{CTbook}.
The inequality in step (b) follows from the fact that the proper complex Gaussian distribution is the entropy maximizer, i.e.,
$h({\bf v}_k \, | \, {\bf H})  \leq \log_2 ( (\pi \, e)^T \, \vert {\bf R}_v \vert)$, where ${\bf R}_v \Define {\mathbb E}[ {\bf v}_k {\bf v}_k^H]$ is the autocorrelation matrix of ${\bf v}_k$ and $\vert {\bf R}_v \vert$ denotes its determinant \cite{propercomplex}.
From (\ref{recvk_eqn_tilde}) and the definition of ${\bf v}_k$ in (\ref{inf_rate_bnd}) we get ${\bf v}_k =    - {\bf I}_k^u  \, - \, {\big (} {\bf w}_k/\sqrt{P_T} {\big )}$. Since ${\bf I}_k^u$ and ${\bf w}_k$ are independent,
it follows that ${\bf R}_v \, = \, {\mathbb E} [  {\bf I}_k^u \, {\bf I}_k^{u^H} \, | \, {\bf H} ] \, + \, (\sigma^2/P_T) {\bf I}$, where the expectation is over ${\bf u}_1, \cdots, {\bf u}_M$.
Substituting this expression for ${\bf R}_v$ in (\ref{inf_rate_bnd}), we get
{
\small
\begin{eqnarray}
\label{rk_eqn}
\frac{I({\bf y}_k \, ; \, {\bf u}_k \,\, | \,\,  {\bf H} )}{T } & \geq &  R_k({\bf H},{\bf E},\frac{P_T}{\sigma^2})  \,\,,\,\, \mbox{where}  \nonumber \\
\hspace{-1mm} R_k({\bf H},{\bf E},\frac{P_T}{\sigma^2}) &  \hspace{-5mm} \Define &  \hspace{-6mm} {\Bigg [ }
\log_2(E_k)    -  \frac{\log_2{\Big \vert}  {\mathbb E} [  {\bf I}_k^u \, {\bf I}_k^{u^H} \, | \, {\bf H} ] \, + \, \frac{\sigma^2}{P_T} {\bf I}     {\Big \vert} }{T }  {\Bigg ] }^+
\end{eqnarray}
\normalsize
}
Here $[ x ]^+ \, \Define \, \max( 0 , x)$ and ${\bf E} \Define (E_1,E_2,\cdots,E_M)^T$ is the vector of the average information symbol energies of the $M$ users.
%Note that $R_k({\bf H},{\bf E},P_T/\sigma^2)$) is an achievable rate for the $k$-th user.
The ergodic information rate lower bound for the $k$-th user is then given by ${\mathbb E}[ R_k({\bf H},{\bf E},P_T/\sigma^2) ]$ (expectation is over ${\bf H}$).\footnote{\footnotesize{
In this paper, $E$ is fixed and does not vary with $H$. The achievable sum rate could
be improved by adapting $E$ with ${\bf H}$, but then this would be difficult to realize in practice since we do not know
the exact analytical dependence of the optimal $E$ (which maximizes the ergodic sum rate) on the instantaneous ${\bf H}$.}}
\section{Numerical results and discussion}
\label{sec-sim}
We consider a frequency selective channel with a uniform power delay profile, i.e., ${\mathbb E}[\vert h_{k,i}[l] \vert^2] = 1/L \,,\,l=0,1,\cdots,(L-1)$.
The channel gains $h_{k,i}[l]$ are i.i.d. Rayleigh faded, i.e., proper complex Gaussian (mean $0$, variance $1/L$).
The ergodic sum rate $\sum_{k=1}^M {\mathbb E}[ R_k({\bf H},{\bf E},P_T/\sigma^2) ]$ can be maximized as a
function of $(E_1,\cdots,E_M)$. This is however difficult. Nevertheless, since the users have identical channel
statistics, it is likely that the optimal ${\bf E}$ vector has equal components, i.e., $E_k={E}^{\prime} \,,\, k=1,\ldots,M$.\footnote{\footnotesize{Since $E$ does not vary with each channel realization ${\bf H}$, the optimal $E$ depends on the multi-user channel only through its statistics.}}
Using numerical methods, for a given $P_T/\sigma^2$ we therefore find the optimal ${E}^{\prime}$ which results in the largest ergodic sum rate.
With $E_k={E}^{\prime} \,,\, k=1,\ldots,M$, we observe that all users have the same ergodic rate, i.e. ${\mathbb E}[ R_1({\bf H},{\bf E},P_T/\sigma^2) ] = \cdots = {\mathbb E}[ R_M({\bf H},{\bf E},P_T/\sigma^2) ]$.
Subsequently, we refer to this rate achieved by each user as the per-user ergodic information rate.
\begin{figure}[t]
\vspace{-1mm}
\begin{center}
\epsfig{file=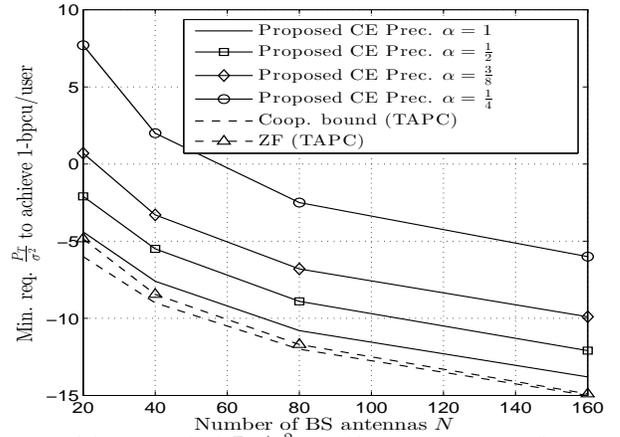, width=80mm,height=58mm}
\end{center}
\vspace{-6mm}
\caption{Minimum required $P_T/\sigma^2$ to achieve a per-user ergodic rate of $1$ bpcu, plotted as a function of increasing $N$. Fixed $M=5$ users and $L=4$.}
\label{fig_3}
\vspace{-6mm}
\end{figure}

In Fig.~\ref{fig_3} we plot the minimum $P_T/\sigma^2$ required by the proposed CE precoder to achieve a per-user information rate of $1$ bpcu
as a function of increasing $N$ with fixed $M=5$ users and $L=4$. The special case of $\alpha=1$ corresponds to an unconstrained (time-variation) CE precoder and therefore has the
best performance. We see that for a given $N$, more transmit power is required for a smaller $\alpha$. This is expected since a smaller $\alpha$ places a more stringent constraint on the time-variation of the transmitted phase angles, which reduces the information rate.
However, even with $\alpha = 1/2$ (i.e., limiting the magnitude of the time variation between consecutive time instances
to be less than $90^{\circ}$), the extra transmit power required when compared to $\alpha=1$ is less than $2$ dB when $N$ is sufficiently larger than $M$ (in this case $N > 8 M$). Also, for a fixed $\alpha$ the extra transmit power required when compared to $\alpha=1$, decreases with increasing $N$. From the figure, it is also observed that irrespective of the value of $\alpha$, for sufficiently large $N \gg M$ the required $P_T/\sigma^2$ reduces by roughly $3$ dB with every doubling in $N$ (i.e., an $O(N)$ array gain with $N$ BS antennas). For the sake of completeness, we have also considered the sum rate achieved under only an average total transmit power constraint (TAPC) which is clearly less stringent than the per-antenna CE constraint. Under TAPC, we have plotted an achievable sum rate (ZF - Zero-Forcing precoder) and an upper bound on the sum capacity (cooperative users). It can be observed that even with $\alpha=1/2$, the extra total transmit power required by the CE precoder when compared to the sum capacity achieving precoder under TAPC, is roughly $3$ dB when $N \gg M$. CE precoding with non-linear PAs is beneficial, since this $3$ dB loss is less than the gain in power efficiency that one can achieve by using a non-linear power-efficient PA instead of using a highly linear inefficient PA \cite{cripps}.


\begin{thebibliography}{1}
\bibitem{SPM-paper}
F.\ Rusek, D.\ Persson, B.\ K.\ Lau, E.\ G.\ Larsson, O.\ Edfors, F.\ Tufvesson and T.\ L.\ Marzetta,
``Scaling up MIMO: opportunities and challenges with very large arrays,''
{\em IEEE Signal Process. Mag.}, vol.\ 30, no.\ 1, pp.\ 40-46, Jan. 2013.
\bibitem{NextGen}
E.\ G.\ Larsson, O.\ Edfors, F.\ Tufvesson and T.\ L.\ Marzetta, ``Massive MIMO for next generation wireless systems,''
 {\em IEEE Commun. Mag.}, vol.\ 52, no.\ 2, pp.\ 186-195, Feb. 2014. 
\bibitem{TM}
T.\ L.\ Marzetta, ``Noncooperative cellular wireless with unlimited number of base station antennas,'' {\em IEEE Trans. Wireless Commun.}, vol.\ 9, no.\ 11, pp.\ 3590-3600, Nov. 2010.
\bibitem{HNQ}
H.\ Q.\ Ngo, E.\ G.\ Larsson and T.\ L.\ Marzetta, ``Energy and spectral efficiency of very large multi-user MIMO systems,''
{\em IEEE Trans. Commun.}, vol.\ 61, no.\ 4, April 2013.
\bibitem{SAIF}
S.\ K.\ Mohammed, ``Impact of transceiver power consumption on the energy efficiency of zero-forcing detector in massive MIMO
systems,'' to appear in {\em IEEE Trans. Commun.}, 2014.   
\bibitem{cripps}
S.\ C.\ Cripps, \emph{RF Power Amplifiers for Wireless Communications,} Artech Publishing House, 1999.
\bibitem{prev_work}
S.\ K.\ Mohammed and E.\ G.\ Larsson, ``Per-antenna constant envelope precoding for large multi-user MIMO systems,''
{\em IEEE Trans. Commun.}, vol.\ 61, no.\ 3, pp.\ 1059-1071, March 2013.
\bibitem{prev_work2}
S.\ K.\ Mohammed and E.\ G.\ Larsson, ``Constant-envelope multi-user precoding for frequency-selective massive MIMO systems,''
 {\em IEEE Wireless Communications Letters}, vol.\ 2, no.\ 5, pp.\ 547-550, October 2013.  
\bibitem{CTbook}
T.\ M.\ Cover, \emph{Elements of Information Theory,}
{\em John Wiley and Sons}, second edition, 2006.
\bibitem{propercomplex}
F.\ D.\ Nesser, J.\ L.\ Massey, ``Proper complex random processes with applications to information theory,''
{\em IEEE Trans. Info. Theory}, Vol.\ 39, no.\ 4, July 1993.

\end{thebibliography}
\end{document}